\documentclass[epj]{svjour}
\onecolumn
\usepackage{latexsym}
\usepackage{graphics}
\usepackage{comment}
\usepackage{amssymb}
\usepackage{multirow}
\usepackage{ulem}
\usepackage{lineno}
\usepackage{soul,graphicx,xcolor}
\usepackage{epsf} 
\usepackage{slashed}
\usepackage{bm}
\usepackage[intlimits]{amsmath}
\usepackage{amsfonts}
\usepackage{dsfont}
\usepackage{subfigure}
\usepackage{colortbl}
\usepackage{array}
\usepackage{slashed}
\usepackage{wasysym}
\usepackage{graphicx}
\usepackage{dcolumn}
\usepackage{units}
\usepackage[numbers,sort&compress]{natbib}

\newcommand{\epem}{l^+l^-}
\newcommand{\auau}{$\textrm{Au}$+$\textrm{Au}$}
\newcommand{\inin}{$\textrm{In}$+$\textrm{In}$}

\newcommand{\pdA}{$\textrm{p}/\textrm{d}$+$\textrm{A}$}
\newcommand{\agev}{$A$~GeV}
\newcommand{\gevcc}{GeV/$c^{2}$}

\newcommand{\gev}{GeV}
\newcommand{\mev}{MeV}
\newcommand{\fmc}{fm/$c$}
\def\etal{{\it et al.\/}}

\newcommand{\eg}{{\it e.g.}}
\newcommand{\ie}{{\it i.e.}}
\newcommand{\dup}{\mathrm{d}}
\definecolor{webgreen}{rgb}{0,0.75,0}
\definecolor{webred}{rgb}{0.75,0,0}
\definecolor{webblue}{rgb}{0,0,0.75}
\definecolor{darkblue}{rgb}{0,0,0.6}
\definecolor{dunkelgrau}{rgb}{0.8,0.8,0.8}
\definecolor{lgray}{rgb}{0.95,0.95,0.95}
\definecolor{lgreen}{rgb}{0.95,1.00,0.90}
\definecolor{lblue}{rgb}{0.9,0.95,1.00}
\definecolor{lred}{rgb}{1.00,0.92,0.85}
\definecolor{shadecolor}{rgb}{1.00,0.92,0.82}
\usepackage[colorlinks=true,linkcolor=darkblue ,citecolor=green,urlcolor=darkblue]{hyperref}
\begin{document}
%\linenumbers
%============================================================
%
\title{Thermal Dileptons from Coarse-Grained Transport as Fireball Probes at SIS Energies}
%\subtitle{}
\author{Tetyana Galatyuk\inst{1,2}, Paul M. Hohler\inst{3}, Ralf Rapp\inst{3}, 
Florian Seck\inst{1,2} \and Joachim Stroth\inst{3,2}
}
%
%\offprints{f.seck@gsi.de}% Insert a name or remove this line
\mail{f.seck@gsi.de}
\institute{Technische Universit\"{a}t Darmstadt, 64289~Darmstadt, Germany
      \and GSI Helmholtzzentrum f\"{u}r Schwerionenforschung GmbH, 64291~Darmstadt, Germany
      \and Texas A$\&$M University, College Station, TX 77843-3366, USA
      \and Institut f\"{u}r Kernphysik, Goethe-Universit\"{a}t, 60438 Frankfurt, Germany}
 \date{} %Received: 29.12.2015 / Revised version: }
\abstract{
Utilizing a coarse-graining method to convert hadronic transport simulations of 
\auau\ collisions at SIS energies into local temperature,  baryon and pion
densities, we compute the pertinent radiation of thermal dileptons based on
an in-medium $\rho$ spectral function that describes available spectra at 
ultrarelativistic collision energies. In particular, we analyze how far
the resulting yields and slopes of the invariant-mass spectra can probe the 
lifetime and temperatures of the fireball.
We find that dilepton radiation sets in after the initial overlap phase 
of the colliding nuclei of about 7~\fmc, and lasts for about 13~\fmc. This duration 
closely coincides with the development of the transverse collectivity of the baryons, 
thus establishing a direct correlation between hadronic collective effects and thermal 
EM radiation, and supporting a near local equilibration of the system. This
fireball ``lifetime" is substantially smaller than the typical 20-30~\fmc\ that naive 
considerations of the density evolution alone would suggest. We furthermore find 
that the total dilepton yield radiated into the invariant-mass window of 
$M=0.3-0.7$~\gevcc\, normalized to the number of charged pions, follows a 
relation to the lifetime found earlier in the (ultra-) relativistic regime of 
heavy-ion collisions, and thus corroborates the versatility of this tool.
The spectral slopes of the invariant-mass spectra above the $\phi$-meson mass
provide a thermometer of the hottest phases of the collision, and agree well 
with the maximal temperatures extracted from the coarse-grained hadron spectra.
 % end of PACS codes
} %end of abstract
\authorrunning{T. Galatyuk, P.M. Hohler, R. Rapp, F. Seck, J. Stroth}
\titlerunning{Thermal dileptons as fireball probes at SIS18}
\maketitle
%
%============================================================
\section{Introduction}
\label{intro}
%============================================================
Lepton pairs have proven to be a formidable tool to probe the properties of extreme 
states of QCD matter as formed in reactions of heavy ions at (ultra-) relativistic 
energies~\cite{Rapp:2009yu}. Since electromagnetic (EM) probes decouple from 
the hot and dense interaction region once they are produced, their phase space 
distributions carry information about the temperature, collectivity and spectral 
structure of the QCD medium~\cite{Rapp:2014hha}. The recorded spectra are a 
space-time integral of a local, time-dependent emissivity of matter over the 
full reaction volume. The emissivity characterizes the radiation rate of (virtual)
photons off a cell of strongly interacting matter per unit time and 4-momentum,
$\epsilon= dN_{\rm ee}/dVdtd^4q$. 
In thermal equilibrium it depends on intensive medium properties such as temperature, 
density and chemical composition, and can be represented by the thermal average of 
the in-medium EM current-current correlator~\cite{Pisarski:1982,McLerran:1984ay,Gale:1990pn},
$\epsilon = {\rm const} f^B(q_0,T) \rho_{\rm EM} /M^2$, where $M$ is the invariant 
dilepton mass and $f^B$ the thermal Bose distribution. 
In the vacuum, and at low masses, the EM spectral function $\rho_{\rm EM}$, is 
saturated by the decays of the light vector 
mesons $\rho$, $\omega$ and $\phi$, while at very high temperatures (or high masses) 
it is characterized by the annihilation of weakly interacting quarks and antiquarks. 
A key objective of using dilepton emission is to study the modifications of hadron 
properties in a QCD medium and how these can signal the transition to deconfined 
and/or chirally restored phases of matter~\cite{Rapp:2009yu,Leupold:2009kz,HohlerRapp2014}.

An important aspect in understanding dilepton radiation off matter under extreme 
conditions is the excitation function as the energy of the colliding nuclei is varied. 
On the one hand, this allows to vary the chemical composition of the system and to scan
how hadron properties change across the QCD phase diagram, \ie, to study the microphysics
encoded in the EM emissivity. On the other hand, if one has control over the EM 
spectral function, one can utilize dileptons to study the macrophysics of the fireball, 
as their spectra are determined by an interplay of the fireball's lifetime and volume
with the strong (exponential) temperature dependence in the Bose distribution of the 
emissivity. Long lifetimes and large fireball volumes are typically associated with the
later emission stages while a high temperature enhances the yields from the early
phases.  
To study this interplay over the full range of collision energies, special care has to 
be taken in the  modeling of the evolution of the reaction volume. At ultrarelativistic 
collision energies (CERN/SPS, BNL/RHIC and CERN/LHC) the source is believed to be close
to local equilibrium and produce most of its entropy very quickly (within about 1~\fmc),
forming a system of extreme energy density in the deconfined phase. It subsequently cools 
through rapid expansion and, after $\tau_{\rm QGP}\simeq5-10$~\fmc, crosses over to a 
hadronic medium at a temperature of about 160~\mev, followed by a further expansion of 
$\tau_{\rm had}\simeq5-10$~\fmc\ until thermal freeze-out at $T_{\rm fo}\simeq 100-120$~\mev. 
In this energy regime the observed dilepton 
spectra~\cite{Agakichiev:2005ai,na60,starBES,star200gev,star200gev_long,phenix200gev_new}
can be understood using isentropically expanding fireballs or hydrodynamic models
for the bulk medium 
evolution~\cite{vanHees:2007th,RappBES_LHC,Renk_sps,Dusling_sps,Vujanovic,Ruppert_sps}.  

At the relativistic energies of LBNL/BEVALAC and GSI/SIS18 the situation is different. 
The Lorentz contraction of the incoming nuclei is moderate, and it already takes around 
7~\fmc\ for two heavy ions (Au) to fully penetrate. Microscopic transport calculations 
predict that the maximum density in the center of the interaction region of a central 
\auau\ collision at 1\agev\ lab energy is reached about 10~\fmc\ after initial 
impact~\cite{Friman:2011zz}, and that the total fireball lifetime is around
25~\fmc. A long-standing question in this context is whether local thermalization
occurs in these systems, and, if so, over which period in time~\cite{Belkacem:1998, Lang:1991}.
Thus, in lieu of thermal approaches, dilepton spectra at these 
energies~\cite{dls1989,dls1997,hades_arkcl,hades_auau} are commonly calculated
using non-equilibrium transport 
models~\cite{bass_urqmd,gibuu,hsd,Wolf:1990,Wolf:1993,Aichelin:2007,linnyk_ko_phsd,greiner,santini}. 
However, in view of the expected strong medium effects on the hadronic spectral
functions, a full quantum treatment of the associated off-shell effects is challenging. 
From the experimental side, one can monitor the dilepton excess radiation 
beyond final-state decays as a function of nucleon participant number, $N_{\rm part}$; 
an excess yield scaling stronger than linear in $N_{\rm part}$ can serve as a measure 
of the number of $\Delta$ or $\rho$ generations produced during the fireball evolution 
at SIS (``$\Delta$-clock")~\cite{hades_auau} or SPS energies 
(``$\rho$-clock")~\cite{damjanovic,na60_2010}, respectively. However, medium effects 
on their spectral functions compromise this measure. A quantitative measure of the 
fireball lifetime has recently been put forward in the ultrarelativistic 
regime~\cite{Rapp:2014hha}: the dilepton excess radiation in the low-mass window of 
$M=0.3-0.7$~\gevcc\ turns out to accurately reflect the underlying lifetime of the 
thermal system, if medium effects in the $\rho$ spectral function are included. 

In the present paper we pursue similar ideas to probe fireball properties in the 
relativistic collision energy regime, using, however, different methods due 
to the above mentioned complications in defining a thermal system. Specifically,
we will compute the time dependent emission of dileptons by applying a 
coarse-graining method to an underlying transport evolution, as first carried 
out in Ref.~\cite{Huovinen:2002im} at SPS energies and in recent work in 
Refs.~\cite{Endres:2014,Endres:2015}). The basic assumption of this procedure is 
that the interactions in the system are strong enough so that, by averaging over 
many transport events of the same collision configuration, one can extract meaningful 
local temperatures and baryon densities in space-time. If this assumption is valid, 
one gains the key advantage of being able to compute dilepton radiation by using 
microscopically calculated equilibrium rates, thus retaining their full quantum-field 
theoretical (off-shell) properties. Our goal in doing so is to identify observables 
which can differentiate between effects of the emissivity and the space-time evolution. 
In particular, we differ from and go beyond the recent work in Ref.~\cite{Endres:2015} 
in several respects. This includes technical aspects such as the extraction of the 
temperature (with quantitative error estimates and their manifestation in the resulting 
dilepton yields) and an improved determination of the pion fugacity factor figuring in 
the dilepton production rate. We furthermore exhibit novel insights, such as a strong 
correlation of the time evolution of dilepton emission with the build-up of collectivity 
in the fireball (which also serves as an independent means to assess the validity of the 
coarse-graining procedure) and quantifying the temperature(s) and lifetime of the fireball.

Our paper is organized as follows.
In Sec.~\ref{sec:coarse} we briefly review the coarse-graining procedure, our concrete
implementation thereof, and specify the in-medium EM emissivity employed in the calculations. 
In Sec.~\ref{sec:fireball} we discuss the extraction and results for the time evolution
of the thermodynamic parameters in Au-Au collisions at SIS18.
In Sec.~\ref{sec:dilep} we analyze the time profile of dilepton radiation, extract a 
fireball lifetime and temperatures from various regions of the invariant-mass spectra, 
and discuss the results in a broader context of heavy-ion collisions at varying energies.
We summarize and conclude in Sec.~\ref{sec:concl}.

%
%============================================================
\section{Coarse Graining of Hadronic Transport}
\label{sec:coarse}
%============================================================
%
Microscopic transport simulations of heavy-ion collisions aim at a description of 
the space-time evolution of the phase-space distributions of all strongly-interacting 
particles involved in the reaction. In the SIS18 energy regime, with lab energies
of up to a few \agev, the calculations are usually performed using hadronic degrees
of freedom. For each heavy-ion collision event, a complete reaction history of
the positions and momenta of hadrons is computed, from the initial encounter of the two 
incoming nuclei until the final state when the interactions have ceased, by
simulating the Boltzmann equation with suitable cross sections. 
Dilepton emission is commonly computed by integrating the dilepton branching ratio 
in each time step for all hadrons and scattering events. This is repeated for
many events in a given event class (\eg, centrality) to obtain the average
dilepton spectrum per event. However, the implementation of in-medium effects, 
in particular the treatment of hadrons with broad spectral distributions, remains 
rather challenging, see, \eg, Refs.~\cite{Bratkovskaya:2007jk,Buss:2011mx,Kampfer:2010}. 
On the other hand, the use of hydrodynamics at relativistic energies also faces significant 
issues, most notably the timescales and justification of thermalization given the 
rather long penetration times of the incoming nuclei~\cite{Ivanov:2005yw}.

As a middle ground between the microscopic transport and macroscopic hydro approach
a coarse-graining procedure has been suggested to compute the EM radiation of
the interacting medium in heavy-ion collisions~\cite{Huovinen:2002im}. By averaging
the pion and baryon distributions in suitable space-time cells over many events
one extracts smooth space-time evolutions of temperature and chemical 
potentials in a given event class, which can be straightforwardly convoluted
with a thermal dilepton emission rate. This combines the virtues of a microscopic
space-time evolution with the benefits of thermal-field theoretically calculated 
in-medium dilepton rates. The price is the assumption of local thermal equilibrium
in the extraction of temperature and chemical potentials, which, however, is 
mitigated compared to a full hydrodynamic simulation since deviations from
the vanishing mean-free-path limit are still kept in the evolution (and, in fact, 
are also inherent in the dilepton rates).

The expression for the event-averaged 4-momentum differential dilepton emission 
spectrum is given as the space-time integral of the emissivity as
\begin{equation}
\left< \frac{\dup N_{\epem}}{\dup y \dup p_{t} \dup M} \right> =
\int \frac{\dup\epsilon_{\epem}}{\dup y\dup p_{t}\dup M}
\left( T(x),\mu_i(x),\vec{v}(x) \right)  
\dup^4{x} 
\end{equation}
where $\epsilon_{\epem}$ denotes the rate of dileptons emitted 
per unit time and volume for a 4-momentum $p=(p_0,\vec{p})$ of the virtual
photon of invariant mass $M=\sqrt{p^2}$; $T$, $\mu_i$ and $\vec{v}$ are the local
temperature, chemical potentials and flow velocity of the medium.
For a coarse-grained reaction volume the above expression reduces to 
a discrete sum
\begin{equation}
\left< \frac{\dup N_{\epem}}{\dup y \dup p_{t} \dup M} \right> =
\left( \Delta x \right)^3 \Delta t \sum_{k,l,m,t_j} 
\frac{\dup \epsilon_{\epem}}{\dup y \dup p_{t} \dup M} 
\left( \left< T \right>_{k,l,m,t_j}, \left<  \mu_i \right>_{k,l,m,t_j}, 
\left<\vec{v} \right>_{k,l,m,t_j} \right)  
\label{rate-2}
\end{equation}
where $k,l,m$ label the cartesian coordinates $x,y,z$ of a cubic spatial cell of 
volume $(\Delta x)^3$, $j$ the time-step of duration $\Delta t$, and $\langle\cdot\rangle$ 
the ensemble average for a given cell.  

The thermal emissivity can be expressed in standard form as 
\begin{equation}
\frac{\dup\epsilon_{l^+ l^-}}{\dup^4p} = 
\frac{\alpha_{EM}^2}{\pi^2 M^2} \, f_B(p_0, T) \,
\rho_{\rm{EM}}(M, p; T,\mu_i)
\label{emiss}
\end{equation}
in terms of the EM spectral function $\rho_{\rm{EM}}=-\mathrm{Im}\Pi_{\rm{EM}}/\pi$ of the 
QCD medium ($\Pi_{\rm EM}$: EM current-current correlation function) and the thermal Bose 
distribution function, $f_B$. 
The chemical potentials, $\mu_i$, in Eq.~\eqref{emiss} refer to the baryonic one 
($i$=$B$), but also to effective meson chemical potentials ($i$=$\pi$, $K$) which build
up in the space-time evolution of the thermal fireball after its chemical freeze-out (where, 
by definition, inelastic collisions cease). An over-population of pions ($\mu_\pi>0$) is 
especially important for $\rho$-meson and dilepton production~\cite{Rapp:1999}, as it 
induces an additional overall fugacity factor 
$z_{\pi}^\kappa = \exp\left(\kappa \mu_{\pi}/T\right)$, in the emissivity, where $\kappa$ 
characterizes the number of pions involved in the $\rho$ (dilepton) production process, 
\eg, $\kappa$=2 for $\pi\pi\to\rho$ or $\kappa$=1 for $\pi N\to\rho N$. We will return 
to this issue in Sec.~\ref{sec:dilep} below.  

In hadronic matter and in the low-mass region, $M\le 1.1$\,GeV, the vector 
dominance model directly relates the EM spectral function to the spectral functions 
of the light vector mesons, \eg,
 $\mathrm{Im}\Pi_{\rm EM} =  ({m_{\rho}^4}/{g_{\rho}^2}) \; \mathrm{Im}D_{\rho}$ 
for the dominant $\rho$-meson contribution.
We account for the medium effects on the EM spectral function by employing a 
recently developed parameterization\footnote{The parametrization is available from  
one of the authors (RR) upon request. It is the same one as used in 
Refs.~\cite{Endres:2014,Endres:2015}.} 
of the $\rho$-meson spectral function calculated 
from hadronic many-body theory~\cite{Rapp:1999}. This spectral function is characterized 
by a strong broadening and ultimate ``melting" of the $\rho$ resonance in hot and dense 
hadronic matter; it provides a good description of all available low-mass dilepton data 
in ultrarelativistic heavy-ion collisions~\cite{RappBES_LHC} (including the recently
revised PHENIX data~\cite{phenix200gev_new}), and of nuclear 
photo-production experiments~\cite{Wood:2008ee,Riek:2010gz}. 
For practical use, the parameterization is provided in terms of the dilepton's mass
($M$) and 3-momentum ($p$), and as a function of temperature, pion and kaon
chemical potentials as well as an effective baryon density defined as 
$\varrho_{\textrm{eff}} = \varrho_{N} + \varrho_{\bar{N}} + 
\nicefrac{1}{2}(\varrho_{R} + \varrho_{\bar{R}})$. Here $\varrho_{N(\bar{N})}$ denotes the
density of (anti-) nucleons and $\varrho_{R(\bar{R})}$ of (anti-) baryon resonances 
(including both $N^*$ and $\Delta$ states). The factor of $\nicefrac{1}{2}$ for the latter 
is a conservative estimate, reflecting the finding that $\rho$-induced resonances 
on excited baryons are usually more weakly coupled (and/or less known) than for $\rho N$ 
scattering~\cite{Rapp:1999}. The contributions from anti-baryons are irrelevant in 
the SIS18 energy regime.
This parameterization is generally accurate within a few tens of percent (better for 
space-time integrated spectra), which is sufficient for the purpose at hand. It has 
recently been deployed into a coarse-grained approach at SIS and SPS energies, 
resulting in good agreement with the measurements of HADES in Ar+KCl collisions and 
the high-precision NA60 data~\cite{Endres:2014, Endres:2015} (which, in turn, also
supports the viability of the coarse-graining procedure).
We also include radiation from a hadronic continuum relevant for the intermediate-mass
region (IMR; $M>1$~\gevcc) using a standard parameterization deduced from the cross 
section for \mbox{$e^+e^-\to {\rm hadrons}$} annihilation~\cite{Shuryak:1993kg}, 
\begin{equation}
\rho_{\rm EM}^{\rm cont} (M) = \frac{1}{4\pi^2}\left(1+\frac{\alpha_s}{\pi}\right)
\frac{M^2}{1+\exp[(E_0-M)/\delta]} \ .
\label{rho_cont}
\end{equation}
with a ``threshold" energy $E_0$=1.5\,GeV and width $\delta$=0.2\,GeV, and $\alpha_s=$0.5.
We here neglect medium effects, \eg, due to chiral mixing, which are operative for 
$M<1.5$\,GeV, mostly because they are not material for our present purpose and because 
their assessment in baryon-rich matter is not straightforward.

%
%============================================================
\section{Fireball Evolution}
\label{sec:fireball}
%============================================================
%
For the bulk evolution transport model we employ Ultra-relativistic Quantum Molecular 
Dynamics (UrQMD)~\cite{bass_urqmd} which successfully describes most of the hadron
data in heavy-ion collisions at relativistic energies. We perform the coarse-graining
following the approach of Ref.~\cite{Huovinen:2002im} by simulating an ensemble of 
heavy-ion collisions at fixed centrality at SIS18 energy.
To optimize the conditions for the main assumption underlying the coarse-graining, we
focus on the largest system available, \ie, central \auau\ collisions at 1.23\agev\ \footnote{This 
particular beam energy has been chosen in view of the upcoming HADES results on Au+Au 
reactions taken at this energy.} beam
energy on a stationary target, corresponding to $\sqrt{s_{NN}}=2.4$~\gev. We discretize
the spatial volume into $21^3$ cubic cells of volume $\Delta x \Delta y \Delta z= 1$\,fm$^3$
(covering 10.5~fm in each direction from the center) and analyze them in time steps of 
$\Delta t=1$~\fmc. For each cell we determine an ensemble average 
over many events for the spatial momentum components $p_x$, $p_y$ and $p_z$ of
each particle species, \ie, pions, nucleons and $\Delta(1232)$. 
For the following discussion we divide the cells into two classes. We define an {\it inner 
cube} of cells covering a volume of 7$^3 \times 1$~fm$^3$ around the collision center,
and an {\it outer shell} containing in total $(21^3-7^3)$~fm$^3$. 
\begin{figure*}[!t]
   \centering
   \includegraphics[keepaspectratio,width=0.35\textwidth]{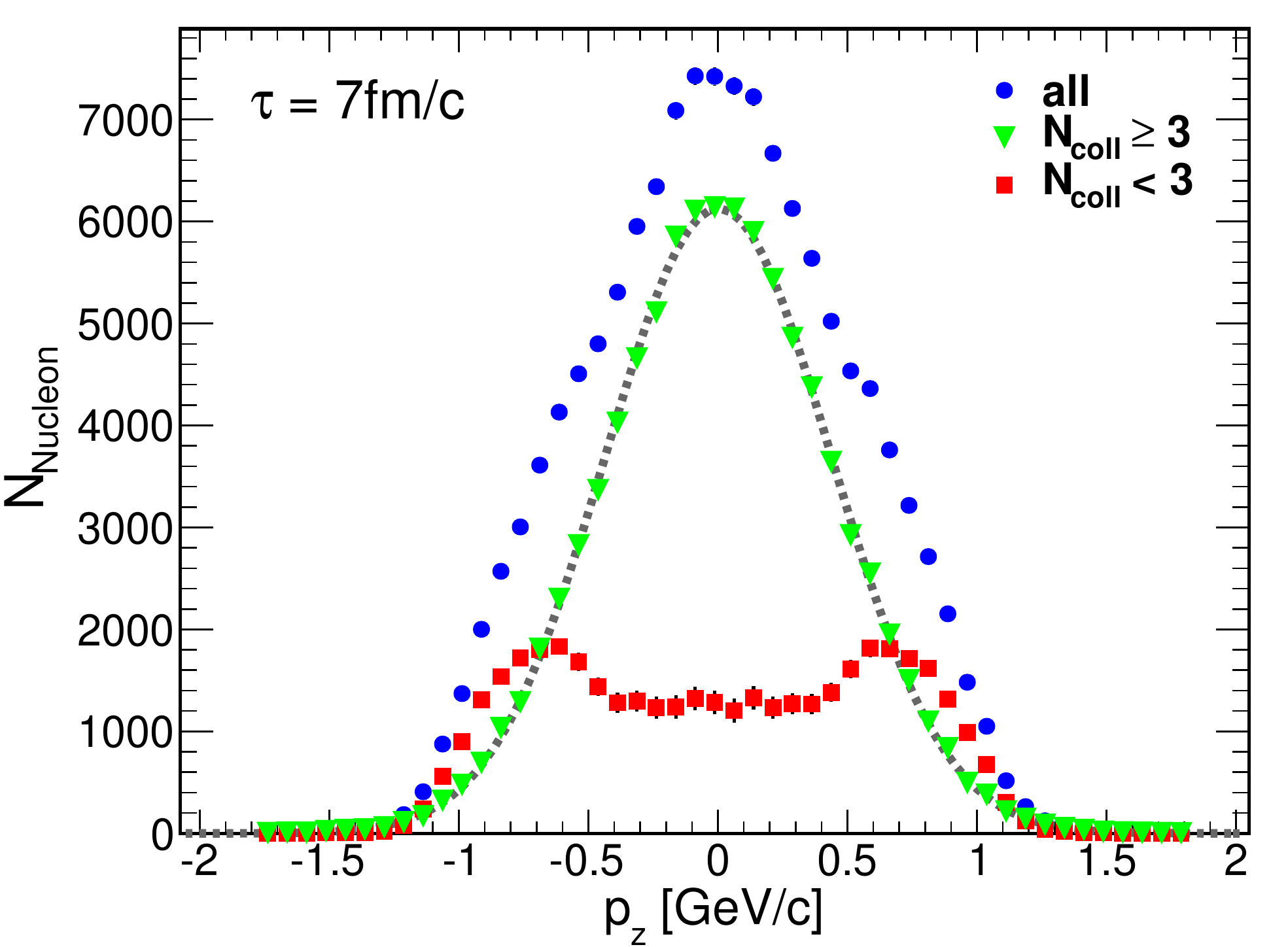}
   \hspace{2cm}
   \includegraphics[keepaspectratio,width=0.35\textwidth]{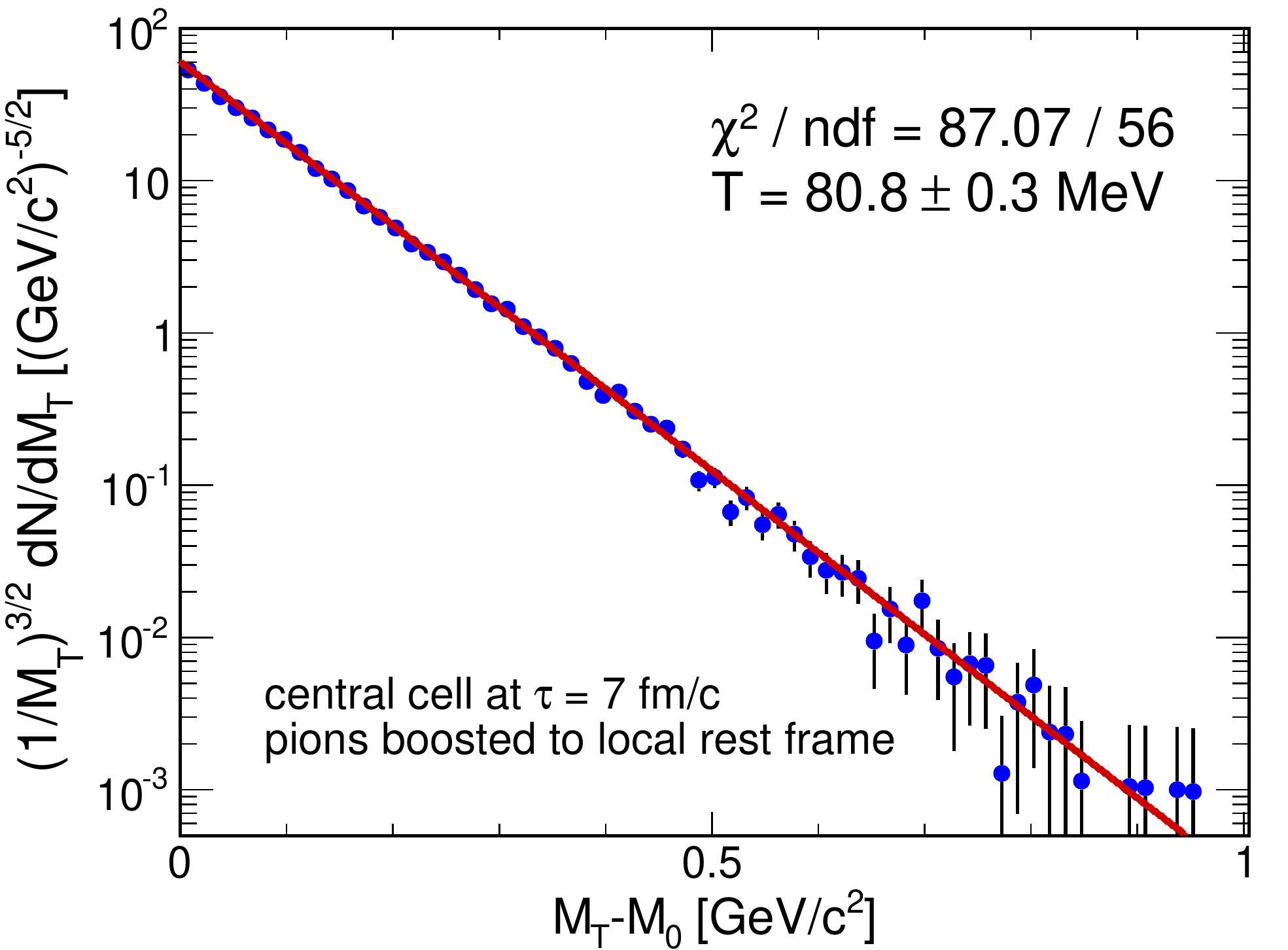}
   \caption{
   (Color online) Left panel: Distribution of the longitudinal momentum of nucleons
   for the time step at 7~\fmc\ after initial nuclear impact in the central cell. 
   The blue circles refer to all nucleons, while the red squares only include nucleons
   which have experienced less than three collisions. Nucleons which underwent 
   three or more interactions are represented by the green triangles. The gray dotted
   line shows a Gaussian fit to this contribution.
   Right panel: $M_T$ spectrum of pions in the central cell at the same time step at
   7~\fmc\ (blue circles). The red line is an exponential fit to UrQMD ``data"
   to extract a temperature.
   }
\label{fig:pz}
\end{figure*}

We first illustrate how the longitudinal momentum distribution of the incoming
nucleons, carrying the beam momentum, changes in the early stages between first
impact and full overlap. In the center of the collision, after approx.~3~\fmc, a 
Gaussian distribution around $p_z=0$ starts to build up from the nucleons which
have collided three times or more. After 7~\fmc, this Gaussian component makes up  
$\sim$70\% of the nucleons (see left panel of Fig.~\ref{fig:pz}), indicating a 
remarkably rapid trend toward thermalization.
Another measure to judge the degree of thermalization within the cells is the
transverse-mass distribution for a given particle species, either 
$M_T^{-2} \, dN/dM_T$ at midrapidity or $M_T^{-3/2} \, dN/dM_T$ when integrated over 
all rapidities~\cite{Hagedorn:1965}. In case of thermalization the spectra will exhibit 
an exponential shape.  The right panel of Fig.~\ref{fig:pz} shows that, after the 
nuclear penetration of 7~\fmc, pions are well described by a thermal 
$M_T$ spectrum of temperature $T\simeq 80$\,MeV.
\begin{figure*}[!t]
   \centering
   \hspace{0.62cm}
   \includegraphics[keepaspectratio,width=0.364\textwidth]{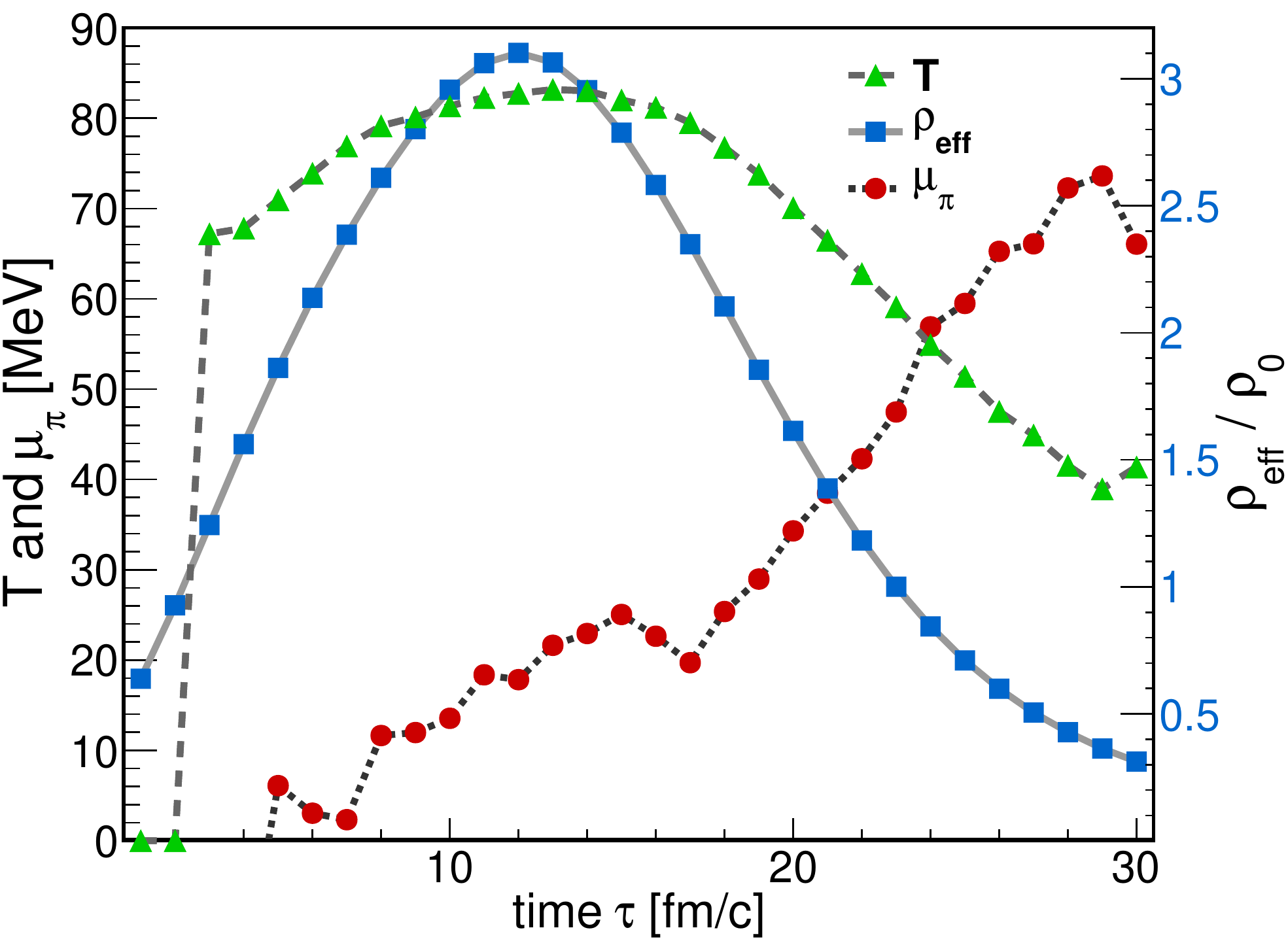}
   \hspace{1.77cm}
   \includegraphics[keepaspectratio,width=0.364\textwidth]{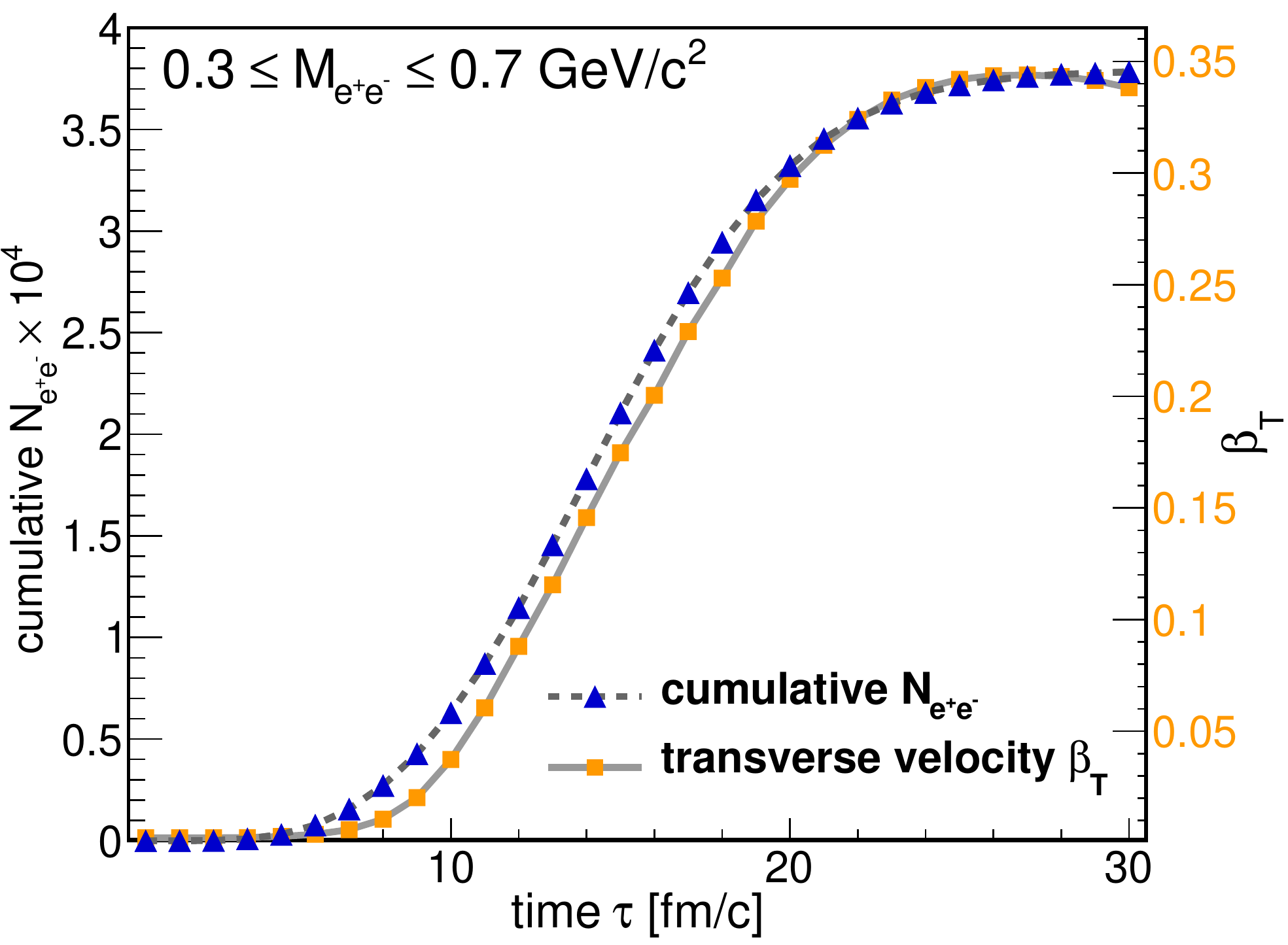}
   \caption{(Color online) Left panel: Time evolution of the ensemble-averaged temperature 
   extracted from pion $M_T$ spectra (green triangles), effective baryon density (blue squares, 
   right vertical scale in units of nuclear saturation density, $\varrho_0$=0.16\,fm$^{-3}$)
   and pion chemical potential (red circles) averaged over the inner cube of $7^3$ cells
   (each of volume 1~fm$^3$) around the collision center. 
   Right panel: Time evolution of the cumulative radiated dilepton yield in the mass range 
   $M=0.3-0.7$~\gevcc\ (blue triangles) and of the average collective transverse velocity in 
   the outer shell of cells (orange squares, right vertical scale).}
\label{fig:time-evo}
\end{figure*}

Next, we extract the local cell temperatures as well as baryon and pion densities defined in 
the rest frame of the cells. We determine the rest frame by evaluating the baryon four-current 
for each cell which yields its collective velocity $\vec v_{klm}(t_j)$. This velocity is 
used to boost all hadron momenta into the cell's rest frame which readily yields its 
baryon (both nucleon and $\Delta$) and pion density $\varrho_{i,\,klm}(t_j)$.
The rest frame momenta of the different particles can then be used to fill properly
normalized $M_T$ spectra (as in the  right panel of Fig.~\ref{fig:pz}). The inverse 
slopes of exponential fits to the distributions of pions are used to define mean local
temperatures $T_{klm}(t_j)$ for all cells and all time steps. Pions are our default 
choice for this purpose since they are strictly produced, the lightest hadrons and hence
abundantly created and the key catalysts for producing dileptons. However, below we will
discuss the uncertainty for dilepton production associated with this choice by also using
nucleons. As it will turn out, the pions (nucleons) yield temperatures which are slightly
lower (larger) than the method of using the energy density from an underlying equation of state
in Ref.~\cite{Endres:2015}. This result is not unexpected (since the EoS method in essence
averages over all species), but it provides an error estimate for this central component 
of the coarse-graining procedure.  With temperature and pion densities at hand, the pion 
chemical potential is determined through a fugacity factor $z_\pi$ in Boltzmann approximation.

For simplicity, we omit the discretization indices in the following. 
The time evolution of the temperature, effective baryon density, 
$\varrho_{\rm eff}=\varrho_N+\frac{1}{2}\varrho_\Delta$, and pion chemical potential,
averaged over the inner cube, are displayed in the left panel of Fig.~\ref{fig:time-evo}.
One finds the well-known result for these collisions energies~\cite{Friman:2011zz} 
that the system evolves for a rather long duration of about 20(25)~\fmc\ 
at baryon densities above (half) normal nuclear matter density. 
Maximal compression is reached after about 10-12~\fmc, and a maximal temperature
of 75-85~\mev\ is maintained for approx.~10~\fmc\ around the maximal compression
phase.
A sizable pion chemical potential develops in the late stages of the fireball evolution.
This is contrary to the coarse-graining results of Ref.~\cite{Endres:2015}, where 
$\mu_\pi$ starts out with a maximum close to maximum compression and then falls to 
zero over time. This discrepancy requires further studies.  
In the right panel of Fig.~\ref{fig:time-evo} we show the transverse collective expansion 
velocity (extracted from the baryon current) as seen by the outer shell of cells. It
starts to develop slightly before maximal compression (but after full nuclear overlap) 
and lasts for a duration of $\sim$12-14~\fmc, until the baryon density has decreased
to near nuclear saturation density.
%
%%%%%%%%%%%%%%%%%%%%%%%%%%%%%%%%%%%%%%%%%
\section{Dilepton Spectra}
\label{sec:dilep}
%%%%%%%%%%%%%%%%%%%%%%%%%%%%%%%%%%%%%%%%%
%
The last ingredient needed to calculate dilepton spectra is the effective fugacity factor, 
$z_{\pi}^{\kappa}$, or more specifically, the effective pion number, $\kappa$, in 
the dilepton production processes at SIS18 energies. At ultrarelativistic energies 
it is routinely taken as $\kappa$=2, as the main production mechanism of the 
$\rho$-meson is via $\pi\pi$ annihilation. In Ref.~\cite{Endres:2015} $\kappa$=2 was
assumed also for SIS18 energies, although the situation for $\rho$ production is 
somewhat different here. To account for this, we have extracted from UrQMD the
individual production channels in order to asses how many pions are involved.
The direct annihilation of two pions contributes $\sim$15$\%$, while the dominant 
channel through baryon resonance decays amounts to $\sim$85$\%$; UrQMD further allows 
to trace back the collision history to determine how many pions were involved in 
creating a resonance that subsequently decays into a $\rho$-meson. We find that 
$\sim$30$\%$ of 
these resonances are produced in $NN$ collisions (0-$\pi$ process), $\sim$50$\%$ with 
1-$\pi$, $\sim$15$\%$ in 2-$\pi$ processes and the remaining $\sim$5$\%$ with three 
or more pions. Thus the average number of pions creating a $\rho$  amounts 
to $\kappa  =  1.12$. We utilize this as our baseline average fugacity factor
(and quantify the consequences when varying it below), in line with the philosophy 
of the coarse graining. For the continuum emission, eq.~(\ref{rho_cont}), the
fugacity factor in a baryon-rich medium is much more difficult to determine (even 
in a meson-dominated medium it is non-trivial~\cite{vanHees:2007th}). As it will turn
out, the radiation in the IMR will be strongly dominated by the hottest phases where
$\mu_\pi$ is small. We therefore neglect the fugacity factor for the continuum
emission and comment on the impact of this approximation below. 

\begin{figure*}[!t]
   \centering
   \includegraphics[keepaspectratio,width=0.35\textwidth]{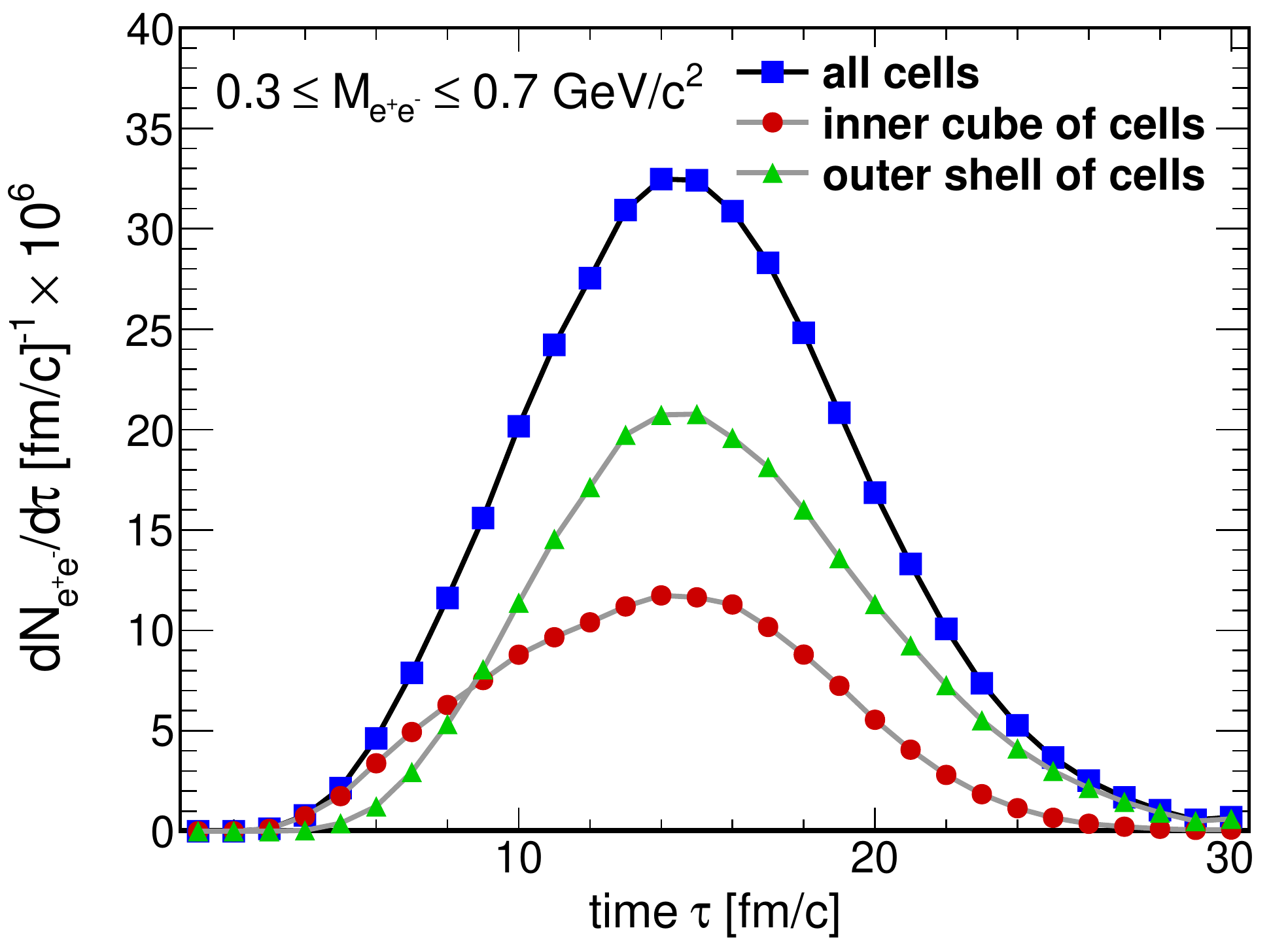}
   \hspace{2cm}
   \includegraphics[keepaspectratio,width=0.35\textwidth]{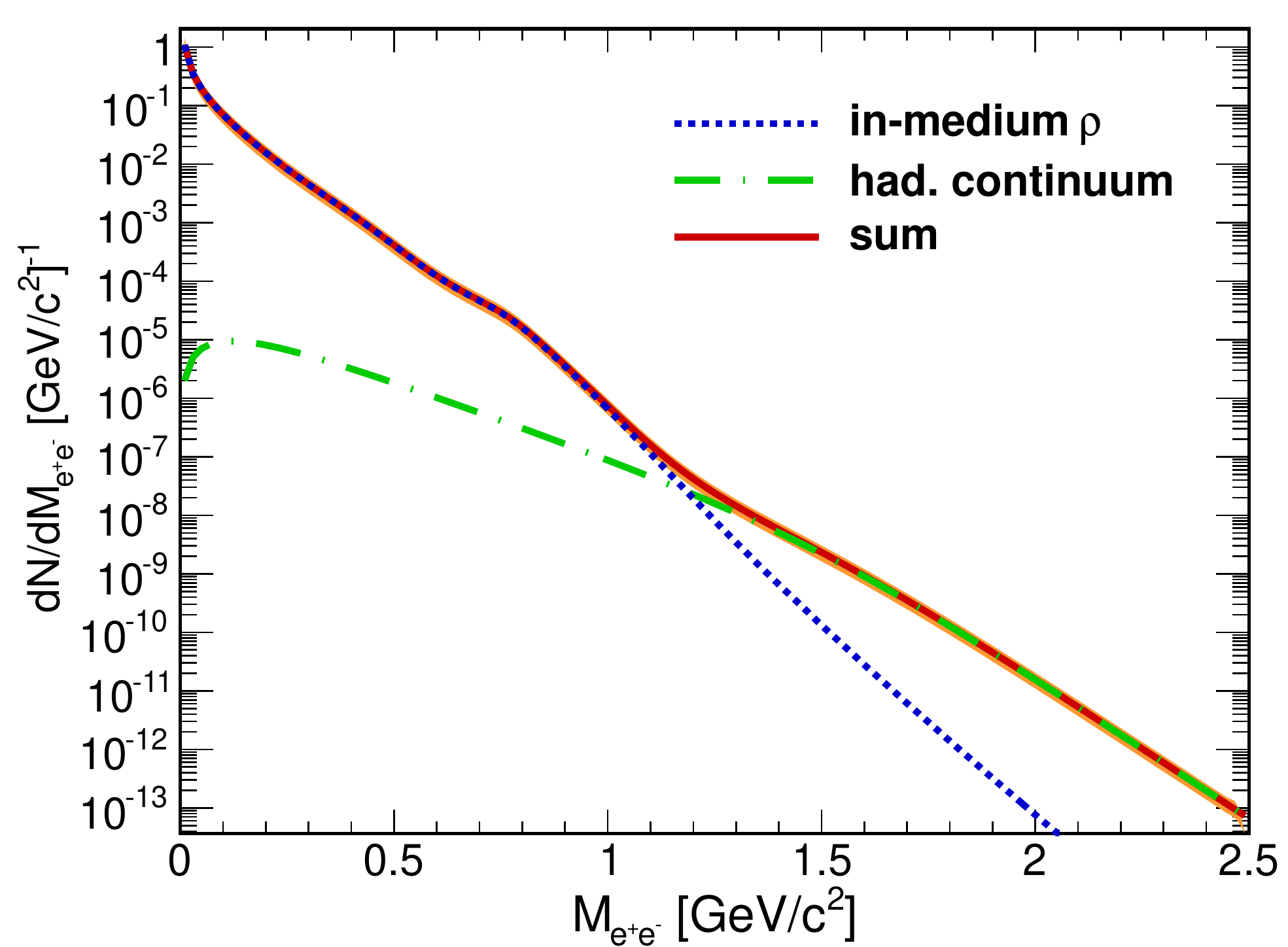}
   \caption{(Color online) Left panel: Time evolution of the radiated dilepton yield per 
   unit time in the mass range \mbox{$M=~0.3-0.7$~\gevcc}. The contribution of all cells 
   together is shown by the blue squares, while the red circles (green triangles) display 
   the share from the inner cube (outer shell) of cells.
   Right panel: Invariant-mass spectrum of $e^+e^-$ pairs radiated from a central \auau\ 
   collision at 1.23\agev. The blue dotted line shows  the contribution of the in-medium 
   $\rho$-meson decays, while the green dotted-dashed lines represents the radiation 
   from the hadronic continuum. The sum of both results in the red solid line which is drawn
   together with an orange error band of $\pm 27\%$.} 
\label{fig:minv_tau}
\end{figure*}

We are now in position to evaluate Eq.~\eqref{rate-2} and to obtain the radiated 
dilepton spectra from the coarse-grained approach, using the local temperatures, 
effective baryon densities and pion chemical potentials in each cell for the emissivity,
Eq.~\eqref{emiss}. To begin with, we integrate (sum up) the 4-momentum differential 
spectrum, Eq.~\eqref{rate-2}, over all \mbox{3-momenta} and over a restricted mass 
window of $M = 0.3 - 0.7$~\gevcc. This window has been identified in a previous 
work~\cite{Rapp:2014hha} to yield a suitable measure of the lifetime of the interacting 
fireball.  The time evolution of the cumulative dilepton yield in this window is displayed
in the right panel of Fig.~\ref{fig:time-evo}. Several interesting features emerge.
First, the active radiation window of $\sim$13~\fmc\ (for $\tau\simeq$~8-21~\fmc\ to 
produce 85$\%$ of the emission) very closely follows the build-up of the collective 
medium flow. Since collectivity is an explicit manifestation of the (thermal) pressure 
in the system, \ie, its thermodynamic response is a consequence of the interactions in 
the system; its clear correlation with ``thermal" dilepton radiation strongly supports
our identification of a ``fireball lifetime".
Second, the radiation duration is quite different from the 20-25~\fmc\ that the inner 
cube spends above nuclear saturation density. In the radiation window, the baryon 
densities and temperatures of the inner cube are above $\varrho_{\rm eff}=1.5\varrho_0$ 
and $T\simeq70$~\mev.\footnote{Note that the 3-volume of the inner cube, with a
``radius" (3.5~fm) of about half the system radius ($R_{\rm Au}\simeq6.5$~fm),
makes up only $\sim$10\% of the total fireball volume. Nevertheless, the integrated
contributions to the dilepton yield from the inner cube ($\sim$40\%) and the outer 
shell ($\sim$60\%) are quite comparable (see left panel of Fig.~\ref{fig:minv_tau}), with 
the former being characterized by a slightly broader emission time distribution than the latter.} 
Third, contributions from the ``pre-equilibrium" phase of nuclear penetration, 
$\tau=$~0-7~\fmc, are negligible, while the EM shining commences shortly thereafter. 
Note that the assumption of thermal equilibrium (``maximum entropy") in our 
calculations provides an upper estimate for the emission from the early phases. At 
the same time this assumption appears to be well satisfied during the radiation 
window (as discussed above), and also leads to a rather rapid and well-defined 
termination of the radiation, suggestive for a thermal freeze-out.  

Next, we test the quantitative relation between fireball lifetime and thermal dilepton 
yield, normalized to the number of charged particles, as put forward in 
Ref.~\cite{Rapp:2014hha} for collider energies. As was done in there, we restrict the 
dilepton and charged-particle numbers to one unit around midrapidity.
Unlike at larger energies, $\sqrt{s_{NN}}\gtrsim5$~\gev, the fireball at lower 
energies is dominated by the incoming nucleons, not by the produced particles. In the 
low-energy limit, the number of charged particles (protons) is therefore not a good 
proxy for the thermal excitation energy in the system. Instead, we normalize the 
dilepton yield to the number of charged pions. With $N_{\epem}=2.9\cdot10^{-4}$ and 
$N_{\pi^\pm}=17.1$ in central \auau\ collisions at $E_{\rm lab}=1.23$\agev, 
one obtains $N_{\epem}/N_{\pi^\pm}=17.0\cdot 10^{-6}$. Based on Fig.~3 in 
Ref.~\cite{Rapp:2014hha}, this translates into a fireball lifetime of 
$\tau_{\rm fb} \simeq 14$~\fmc. On the other hand, if we renormalize the results in 
Fig.~3 of Ref.~\cite{Rapp:2014hha} to the number of charged pions, the proportionality
factor to the lifetime changes to $\sim$1.45, \ie, 
$N_{\epem}/N_{\pi^\pm} \times 10^6\simeq 1.45 \tau_{\rm fb}$ (this includes the
contribution of strong final-state decays, which are also included in the coarse-grained
yields). In this case the extracted lifetime turns out to be $\sim$12~\fmc, not far from 
the $\sim$13~\fmc\ estimated as the ``radiation window".

Let us quantify some of the uncertainties in the coarse-grained approach to calculate
the dilepton yield $N_{e^+e^-}$, as arising from the conversion of the bulk properties
in UrQMD into thermodynamic quantities. Instead of pions, one could use the nucleons
to extract a local temperature from their rest frame $M_T$ (restricted to the ones
which have collided at least three times). This temperature reaches up to 110~\mev\ in the
central cell and is in general up to $30\%$ higher than the temperature extracted
from the pion spectra. However, when using this temperature to determine the corresponding 
pion chemical potentials, much smaller values for the latter emerge (to obtain the same
pion density). Employing the combination of ``nucleon temperature'' and the pertinent 
$\mu_\pi$'s in Eq.~\eqref{emiss} leads to an increase of the dilepton yield, $N_{e^+e^-}$, 
by about $25\%$, which is much smaller than a factor of $\sim$4 which would arise if the 
fugacity factor $z_\pi^\kappa$ were neglected in both calculations. The latter is thus not 
only physically well motivated but also considerably stabilizes our numerical results.
We furthermore examine the uncertainty of the $\kappa$ values which might turn out to be 
somewhat different in transport approaches other than UrQMD. 
When varying $\kappa$ between 1 and 1.3 we find a difference of less than 10\% in the
low-mass dilepton yields compared to our default value.

Finally, we analyze the spectral shape of the calculated invariant-mass spectra of thermal 
dileptons, cf.~right panel of Fig.~\ref{fig:minv_tau} showing their decomposition into 
in-medium $\rho$-meson decays and the hadronic continuum. The strong medium effects on 
the $\rho$-meson lead to a remarkably structureless low-mass spectrum with only a slight 
bump remaining in the region of the vacuum $\rho$ mass. The medium effects appear to be
even stronger than in the NA60 spectrum for \inin\ at 158\agev\ beam energy.
This is in part due to the larger collisions system, the lower temperatures (which
produce steeper slopes), and the increased medium
effects induced by the higher baryon densities. As discussed above, a large
part of the radiation in the 1.23\agev\ \auau\ system emanates from densities above 
2 times saturation density, with a large fraction of nucleons (which, in the 
calculation of the $\rho$ spectral function, induce stronger medium effects than 
excited baryons). This is further augmented by the suppression of the emission from 
the dilute phases, apparently caused by the drop in temperature in the later stages
which cannot be compensated by the increase in volume. In some sense the situation
is similar to the IMR at full SPS energy, where the temperature sensitivity of the
overall Bose factor strongly biases the emission to the early phases.
Since the initial temperatures at SIS are roughly 3 times smaller than at SPS, one 
expects the predominance of early emission roughly at a factor of 3 smaller
invariant masses, which suggests that the role of the $M>1.2$~\gevcc\ region at SPS 
shifts to $M>0.4$~\gevcc\ at SIS18. This is probably the reason for our
finding that the dilute phases contribute little at SIS18.

As is well known, the spectral slope of dilepton invariant-mass spectra is an excellent 
thermometer of the fireball, unaffected by blue-shift effects due to the collective medium
expansion. This is especially true if the EM spectral function does not (or only weakly)
depend on temperature, so that the temperature dependence is solely residing in the 
thermal Bose factor. This is rather well satisfied in the IMR, where the 
vacuum spectral function, Im\,$\Pi_{\rm EM}/M^2$, is essentially a constant 
with thermal corrections suppressed at order $T^2/M^2$.
After integration over 3-momentum, the emissivity is then approximately proportional 
to \mbox{$\dup\epsilon_{\epem}/\dup M \,\propto\,(MT)^{3/2}\,\exp{(-M/T)}$}. 
A fit to our calculated mass spectrum in the IMR for $M=1.5-2.5$~\gevcc\ yields an 
inverse-slope parameter of $T_{\rm s}=88\pm5$~\mev, which coincides with the highest 
temperatures reached in the central cube of the fireball. 
This implies that the pertinent dilepton radiation is entirely dominated by the 
hottest (not earliest!) phases of the fireball evolution, well in line with the 
trend found in the ultrarelativistic regime when lowering the collision energy, 
cf.~Fig.~2 in Ref.~\cite{Rapp:2014hha}.\footnote{We note that if we were to include a 
fugacity factor in the continuum emission, the magnitude of the pertinent dilepton 
yield could increase significantly (depending on the precise value of $\kappa$) 
while not affecting its slope.} 
In addition, facilitated by the strong broadening of the $\rho$-meson in baryonic
matter above saturation density, also the low-mass spectrum in 1.23\agev\
\auau\ collisions exhibits a near-exponential shape. A fit to the mass window  
$M=0.3-0.7$~\gevcc\ yields an inverse slope parameter of $T_{\rm s}=64\pm5$~\mev,
reflecting emission from a broader range of fireball conditions (thus its 
potential to measure the lifetime). Since the exponential sensitivity of
the Bose factor to temperature is reduced for smaller masses, the increasing
volume of the cooling fireball compensates the thermal suppression for a while.  

From the above findings the following picture of thermal dilepton radiation in 
heavy-ion collisions in the few-GeV regime emerges. In the early collision stages, 
the effect of the temperature in the system is compensated by the smallness of the active volume,
\ie, the build-up of entropy is not very fast and, as a consequence, the emitted 
radiation is negligible. This is a rather welcome feature as pre-equilibrium 
radiation is theoretically difficult to assess. We recall that the assumption of 
local equilibrium in the coarse-graining procedure provides an upper limit of the 
emitted radiation, at least for momenta and masses of the order of the temperature. 
Only after the compression (density) has reached about 80$\%$ of its maximum value 
(about 8~\fmc\ after initial impact, shortly after the nuclei have fully overlapped),  
dilepton radiation picks up rather rapidly. At that point, about $80\%$ of the 
nucleons have undergone at least three collisions and have formed a Gaussian momentum 
distribution around midrapidity, suggestive for thermalization. Only about 1~\fmc\ 
thereafter, transverse collectivity develops in response to the earlier created 
thermodynamic pressure, further supporting the notion of local near-equilibrium. 
The radiation phase lasts for around 13~\fmc, until about 21~\fmc. This implies 
that in this regime the drop in temperature is compensated by the growing fireball 
volume, down to temperatures as high as $\sim$65~\mev\ in the central cube (somewhat 
lower in the outer shells). This is independently verified by the inverse-slope 
parameter of the low-mass dilepton spectrum. It further implies that there is 
little dilepton radiation from the stages with baryon densities around nuclear 
saturation density and below. The development of the collective flow also ceases 
in this regime (after all, cold nuclear matter at saturation has vanishing 
pressure).

Finally, we briefly compare our results with the features at ultrarelativistic
collision energies. At top SPS energies UrQMD suggests the formation of densities 
of more than 10~$\rho_0$~\cite{Endres:2014}, with extracted initial temperatures
of up to 250~\mev\ (consistent with the inverse slopes in the intermediate-mass
dilepton spectrum~\cite{Rapp:2014hha} or the naive picture that most of the
entropy has been produced once the nuclei have fully overlapped, after about
$\sim$1~\fmc). Due to the longitudinal Lorentz contraction, the fireball volume 
is quite small in this early stage, producing appreciable dilepton radiation 
mostly in the IMR. Only a small fraction of the low-mass dilepton yield emanates 
from this phase; most of the low-mass yield is radiated throughout the fireball 
evolution with effective baryon densities around $\rho_0$ (after all, the low-mass 
yield probes the lifetime of the fireball).
On the other hand, in the SIS energy regime low-mass radiation is largely emitted
during the time interval in which the highest baryon densities are reached, 
well above $\rho_0$. Thus the low-mass spectrum is more sensitive to the 
baryon driven in-medium modifications. The predicted almost complete flattening 
of the $\rho$ spectral function leads to an almost exponential shape of the low-mass 
spectrum (recall the right panel of Fig.~\ref{fig:minv_tau}), which provides an 
additional thermometer characterizing the radiation-active medium in the few-GeV 
collision energy regime. 
%
%============================================================
\section{Conclusion and Outlook}
\label{sec:concl}
%============================================================
%
In the present work we have investigated thermal dilepton production in
the SIS18 energy regime using the coarse-graining approach to interface a 
microscopic transport description with (locally) thermal dilepton rates. 
For the former we used the well-established UrQMD model, while for the 
latter we employed (a parameterization of) an in-medium $\rho$ spectral 
function that describes available dilepton data at ultrarelativistic energies. 
While this is not a new approach, we were able to extract some intriguing new
insights, focusing on \auau\ collisions at 1.23\agev\ bombarding energy. 
The radiation from the early pre-equilibrium phases during nuclear penetration 
turns out to be negligible, while the main radiation window lasts for only about 
13~\fmc, quite a bit shorter than expectations based on the density 
evolution. The radiation window remarkably coincides with the build-up of 
transverse collectivity in the fireball, which establishes a strong correlation 
between hadronic and EM activity and supports the notion of a locally
near-thermalized system (justifying the coarse-graining approach a posteriori). 
The correlation between collectivity and EM radiation could prove particularly
useful in disentangling initial-state and collective effects in the current 
debate on \pdA\ collisions at LHC and RHIC.
Furthermore, we have investigated the relation between the system's lifetime and 
the low-mass dilepton excess yield and found it to follow the systematics 
put forward recently for ultrarelativistic collision systems. Finally we have
extracted (blue-shift free) slope parameters and found that the intermediate-mass 
region remains an excellent thermometer for the hottest phase. In addition,
thanks to the strong baryon-driven medium effects which essentially flatten
the $\rho$ line shape, the low-mass spectrum also becomes approximately 
exponential, providing an additional thermometer characterizing the conditions 
under which the bulk of the radiation is emitted. 
Altogether, our studies support the universality of dilepton spectra as a 
versatile probe of QCD matter formed in the fireballs of heavy-ion collisions 
over a broad range of energies.
%
%============================================================
\section{Acknowledgements}
%============================================================
%
This work was supported by the U.S. National Science Foundation under 
grant PHY-1306359, by the Humboldt foundation (Germany),
by the Helmholtz-YIG grant VH-NG-823 at GSI and TU Darmstadt (Germany), and by the 
Hessian Initiative for Excellence (LOEWE) through the Helmholtz International Center 
for FAIR (HIC for FAIR). 

%============================================================

%============================================================

%============================================================

\end{document}